\documentclass{iaus}
\usepackage{graphicx}

\title[The MiMeS Project]{The MiMeS Project: \\Magnetism in Massive Stars}   
\author[G.A. Wade et al.]{G.A. Wade$^1$, E. Alecian$^{1,2}$, D.A. Bohlender$^3$, J.-C. Bouret$^4$, J.H. Grunhut$^1$, H. Henrichs$^5$, C. Neiner$^6$, V. Petit$^7$, N. St. Louis$^8$, M. Auri\`ere$^9$, O. Kochukhov$^{10}$, J. Silvester$^1$, A. ud-Doula$^{11}$\\and the MiMeS Collaboration\thanks{www.physics.queensu.ca/$\sim$wade/mimes}}   
\affiliation{$^1$Royal Military College of Canada, $^2$LESIA, France, $^3$Canadian Astronomy Data Centre,$^4$LAM, France,$^5$Ast. Inst. Amsterdam, Netherlands, $^6$GEPI, France, $^7$Universit\'e Laval, Canada, $^8$Univ. de Montr\'eal, Canada, $^{9}$LAT, France, $^{10}$Uppsala University, Sweden, $^{11}$Morrisville State College, USA}    

\pubyear{2009}
\volume{259}  
\pagerange{100--100}
\date{"YOUR MAILING DATE"  and in revised form ??}
 \jname{Cosmic Magnetic Fields: From Planets,
to Stars and Galaxies} \editors{K.G. Strassmeier, A.G. Kosovichev \&
J.E. Beckman, eds.}
\begin{document}

\maketitle

\begin{abstract}
The Magnetism in Massive Stars (MiMeS) Project is a consensus collaboration among the foremost international researchers
of the physics of hot, massive stars, with the basic aim of understanding the origin, evolution and impact of 
magnetic fields in these objects. The cornerstone of the project is the MiMeS Large Program at the Canada-France-Hawaii
Telescope, which represents a dedication of 640 hours of telescope time from 2008-2012. The MiMeS Large Program will
exploit the unique capabilities of the ESPaDOnS spectropolarimeter to obtain critical missing information about the poorly-studied magnetic 
properties of these important stars, to confront current models and to guide theory.
\end{abstract}

\firstsection 
\section{Introduction}

Massive stars are those stars with initial masses above about 8 times that of the sun, eventually leading to catastrophic explosions in the form of supernovae. These represent the most massive and luminous stellar component of the Universe, and are the crucibles in which the lion's share of the chemical elements are forged. These rapidly-evolving stars drive the chemistry, structure and evolution of galaxies, dominating the ecology of the Universe - not only as supernovae, but also during their entire lifetimes - with far-reaching consequences. 

The magnetic fields of hot, higher-mass stars are qualitatively different from those of cool, low-mass stars (e.g. Wade 2003). They are detected in only a small fraction of stars, and they are structurally much simpler, and frequently much stronger, than the fields of cool stars. Most remarkably, their characteristics show no clear correlation with basic stellar properties such as age, mass or rotation (e.g. Mathys et al. 1997, Kochukhov \& Bagnulo 2006).  The weight of opinion holds that these puzzling characteristics reflect a fundamentally different field origin: that the observed fields are not generated by dynamos, but rather that they are {\em fossil fields} - the slowly-decaying remnants of field accumulated or generated during star formation (e.g. Mestel 1999, Moss 2001, Ferrario \& Wickramasinghe 2006). This relic nature potentially provides us with a powerful and unique capability: to study how magnetic fields evolve throughout the various stages of stellar evolution, and to explore how they influence, and are influenced by, the important structural changes that occur during all phases of stellar evolution, from stellar birth to stellar death. 

Although this fossil paradigm provides a powerful framework for interpreting the magnetic characteristics of higher-mass stars, its physical details are only just beginning to be elaborated (e.g. Braithwaite \& Nordlund 2006, Auriere et al. 2007, Alecian et al. 2008a). In particular, our knowledge of the basic statistical properties of massive star magnetic fields is seriously incomplete. There is a troubling deficit in our understanding of the scope of the influence of fields on massive star evolution, and almost no empirical basis for how fields modify mass loss.

The Magnetism in Massive Stars (MiMeS) Project represents a comprehensive, multidisciplinary strategy by an international team of recognized researchers to address the Òbig questionsÓ related to the complex and puzzling magnetism of massive stars. Recently, MiMeS was awarded "Large Program" status by both Canada and France at the Canada-France-Hawaii Telescope (CFHT), where the Project has been allocated 640 hours of dedicated time with the ESPaDOnS spectropolarimeter from late 2008 through 2012. This commitment of the observatory, its staff, its resources and expertise, allocated as a result of an extensive international expert peer review of many competing proposals, will be used to acquire an immense database of sensitive measurements of the optical spectra and magnetic fields of massive stars, which will be applied to constrain models of the origins of their magnetism, the structure, dynamics and emission properties of their magnetospheres, and the influence of magnetic fields on stellar mass loss and rotation - and ultimately the evolution of massive stars. More specifically, the scientific objectives of the MiMeS Project are: 

\begin{itemize}
\item To identify and model the physical processes responsible for the generation of magnetic fields in massive 
stars; 

\item To observe and model the detailed interaction between magnetic fields and massive star 
winds; 

\item To investigate the role of the magnetic field in modifying the rotational properties of massive 
stars; 

\item To investigate the impact of magnetic fields on massive star evolution,and the connection between magnetic fields of non- 
degenerate massive stars and those of neutron stars and magnetars, with consequential constraints on stellar 
evolution, supernova astrophysics and gamma-ray bursts.  
\end{itemize}

\section{Structure of the Large Program}

\begin{figure}
 \includegraphics[width=13.5cm]{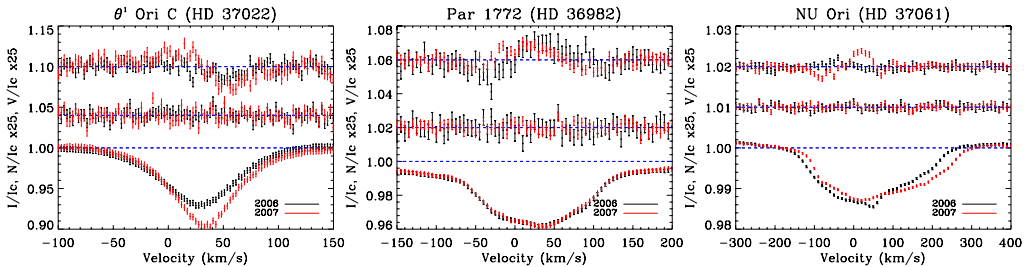}
  \caption{Least-Squares Deconvolved profiles of 3 hot stars: $\theta^1$~Ori C (O7V, left), Par 1772 (B2V, middle) and NU Ori (B0.5V, right). The curves show the mean Stokes $I$ profiles (bottom panel), the mean Stokes $V$ profiles (top panel) and the $N$ diagnostic null profiles (middle panel), black for 2006 January and red for 2007 March. Each star exhibits a clear magnetic signature in Stokes $V$. These results are representative of those expected from the MiMeS Survey Component. From Petit et al. (2008).}\label{fig:wave}
\end{figure}

To address these general problems, we have devised a two-component Large Program (LP) that 
will allow us to obtain basic statistical information about the magnetic properties of the overall 
population of hot, massive stars (the Survey Component), while simultaneously providing detailed 
information about the magnetic fields and related physics of individual objects (the Targeted Component).

{\bf Targeted component:} The MIMeS Targeted Component (TC) will provide data to map the magnetic fields and 
investigate the physical characteristics of a small sample of known magnetic stars of great interest, 
at the highest level of sophistication possible. The roughly 20 TC targets have been selected to allow us to investigate a variety of 
physical phenomena, and to allow us to directly and quantitatively confront models. 

Each TC target will be observed many times using the ESPaDOnS spectropolarimeter, in order to obtain a high-precision and
high-resolution sampling of the rotationally-modulated circular and linear polarisation line profiles. Using state-of-the-art
tomographic reconstruction techniques such as Magnetic Doppler Imaging (Kochukhov \& Piskunov 2002), detailed maps
of the vector magnetic field on and above the surface of the star will be constructed.

{\bf Survey component:} The MiMeS Survey Component (SC) will provide critical missing information about field 
incidence and statistical field properties for a much larger sample of massive stars. It will also serve to 
provide a broader physical context for interpretation of the results of the Targeted Component.  
From a much larger list of potential OB stars compiled from published catalogues, we have 
generated an SC sample of about 150 targets which cover the full range of spectral types from B2-O4 which 
are selected to be best-suited to field detection. Our target list includes pre-main sequence Herbig Be stars,
field and cluster OB stars, Be stars, and Wolf-Rayet stars. 

Each SC target will be observed once or twice during the Project, at very high precision in circular polarisation. From the SC data we will 
measure the bulk incidence of magnetic massive stars, estimate the variation of incidence versus 
mass, derive the statistical properties (intensity and geometry) of the magnetic fields of massive stars, 
estimate the dependence of incidence on age and environment, and derive the general statistical 
relationships between magnetic field characteristics and X-ray emission, wind properties, rotation, 
variability, binarity and surface chemistry diagnostics.  

Of the 640 hours allocated to the MiMeS LP, 385 hours are committed to the SC and 255 hours are committed to the TC. The TC commitment includes 50 hours 
reserved for follow-up of targets detected in the Survey Component.

\begin{figure}
 \includegraphics[width=13.5cm]{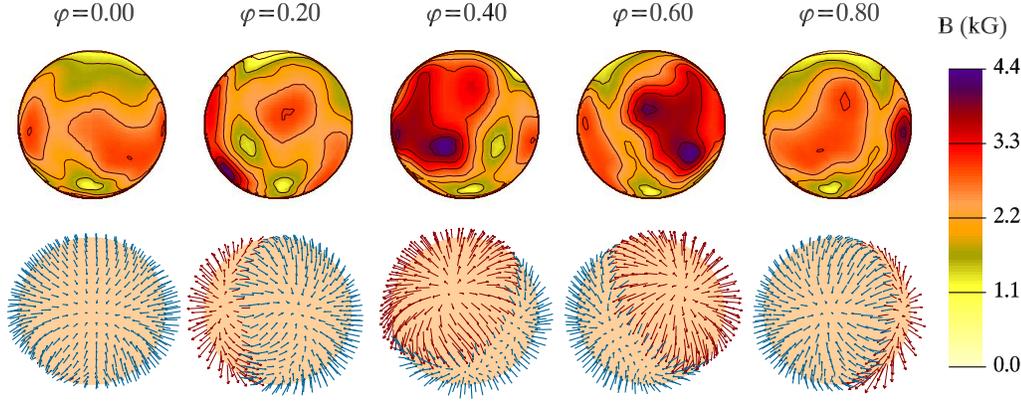}
  \caption{Magnetic Doppler Imaging (MDI) of the B9p star HD 112413 (Kochukhov et al., in preparation), illustrating the reconstructed magnetic field orientation (lower images) and intensity (upper images) of this star at 5 rotation phases. The maps were obtained from a time-series of 21 Stokes $IQUV$ spectral sequences. Although the field line orientation of HD~112413 is approximately dipolar, the field intensity map is far more complex. Maps similar to these will be constructed for the MiMeS Targeted Component.}\label{fig:wave}
\end{figure}

\section{Precision magnetometry of massive stars}

For all targets we will exploit the longitudinal Zeeman effect in metal and helium lines to detect and measure photospheric or pseudo-photospheric magnetic 
fields. Splitting of a spectral line due to a longitudinal magnetic field into oppositely-polarized $\sigma$ components produces a variation of circular polarisation across the line (commonly referred to as a Ò(Stokes $V$) Zeeman signatureÓ or 
Òmagnetic signatureÓ; see Fig. 1.). The amplitude and morphology of the Zeeman signature encode information about the strength and structure of the global magnetic field. 
For some TC targets, we will also exploit the transverse Zeeman effect to constrain the detailed local structure of the field. Splitting of a spectral line by a transverse magnetic field into 
oppositely-polarized $\pi$ and $\sigma$ components produces a variation of linear polarisation (characterized by 
the Stokes $Q$ and $U$ parameters) across the line (e.g. Kochukhov et al. 2004).

\subsection{Survey Component}

For the SC targets, the detection of magnetic field is diagnosed using the Stokes $V$ detection criterion described by Donati et al. (1992, 1997), and the surface field constraint characterised using the powerful Bayesian 
estimation technique of Petit et al. (2008). After reduction of the polarized spectra using the Libre-Esprit optimal extraction code, we employ the Least-Squares Deconvolution (LSD; Donati et al. 1997) multi-line 
analysis procedure to combine the Stokes $V$ Zeeman signatures from many spectral lines into a single high-S/N mean profile (see Fig. 1), enhancing our ability to detect subtle magnetic signatures. Least-Squares 
Deconvolution of a spectrum requires a Òline maskÓ to describe the positions, relative strengths and magnetic sensitivities of the lines predicted to occur in the stellar spectrum. The line mask 
characteristics are sensitive to the parameters describing the stellar atmosphere. In our analysis we employ custom line masks carefully tailored to best reproduce the observated stellar spectrum, in order to maximize the S/N gain of the LSD procedure and therefore 
our sensitivity to weak magnetic fields.

The exposure duration required to detect a Zeeman signature of a given strength 
varies as a function of stellar apparent magnitude, spectral type and projected rotational velocity. This 
results in a large range of detection sensitivities for our targets. The SC exposure times are based on 
an empirical exposure time relation derived from real ESPaDOnS observations of OB stars, and takes into account detection sensitivity gains resulting from LSD and velocity 
binning, and sensitivity losses from line broadening due to rapid rotation.  
Exposure times for our SC targets correspond to the time required to 
definitely detect (with a false alarm probability below $10^{-5}$) the Stokes $V$ Zeeman signature produced 
by a surface dipole magnetic field with a specified polar intensity. Although our calculated exposure times correspond to definite detections of a dipole magnetic field, 
our observations are also sensitive to the presence of substantially more complex field toplogies.

\subsection{Targeted Component}

Zeeman signatures will be detected repeatedly in all spectra of TC targets. The spectropolarimetric timeseries will 
be interpreted using several magnetic field modeling codes at our disposal. For those stars for which 
Stokes $V$ LSD profiles will be the primary model basis, the modeling codes employed by Donati et al. 
(2006) or Alecian et al. (2008b) will be employed.  For those stars for which the signal-to-noise ratio in 
individual spectral lines is sufficient to model the polarisation spectrum directly, we will employ the Invers10 Magnetic Doppler Imaging code to 
simultaneously model the magnetic field, surface abundance structures and pulsation velocity field 
(Piskunov \& Kochukhov 2002, Kochukhov et al. 2004). The resultant magnetic field models will be 
compared directly with the predictions of fossil and dynamo models (e.g. Braithwaite 2006, 2007, Mullan \& Macdonald 2005, Arlt 2008).  

Diagnostics of the wind and magnetosphere (e.g. optical 
emission lines and their linear polarisation, UV line profiles, X-ray photometry and spectroscopy, radio 
flux variations, etc.) will be modeled using both the semi-analytic Rigidly-Rotating Magnetosphere 
approach, the Rigid-Field Hydrodynamics (Townsend et al. 2007) approach and full MHD simulations using the 3D ZEUS 
code (e.g. Stone \& Norman 1992; see Fig. 3). 

\begin{figure}
 \includegraphics[width=13.5cm]{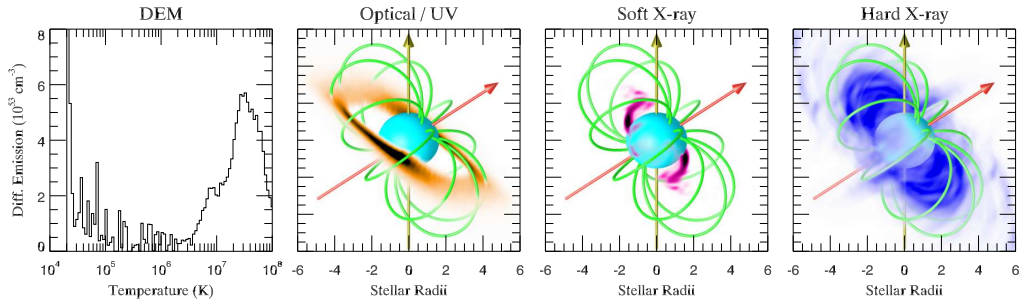}
  \caption{Example of the spectral and spatial emission properties of a rotating massive star magnetosphere modeled using Rigid Field Hydrodynamics (Townsend et al. 2007). The stellar rotation 
axis (vertical arrow) is oblique to the magnetic axis (inclined arrow), leading to a complex potential field produced by radiative acceleration, Lorentz forces and centripetal acceleration. The consequent heated plasma distribution in the stellar magnetosphere (illustrated in colour/grey scale) shows broadband emission, and is highly structured both spatially and spectrally. Magnetically-confined winds such as this are responsible for the X-ray emission and variability properties of some OB stars, and models such as this will be constructed for the MiMeS Targeted Component.
}\label{fig:wave}
\end{figure}

\section{MiMeS data pipeline}

Following their acquisition in Queued Service Observing mode at the CFHT, ESPaDOnS polarised spectra are immediately reduced by CFHT staff using the Libre-Esprit reduction package and downloaded to the dedicated MiMeS Data Archive at the Canadian Astronomy Data Centre in Victoria, Canada. Reduced spectra are carefully normalized to the continuum using existing software tailored to hot stellar spectra. Each reduced ESPaDOnS spectrum is then subject to an immediate quick-look analysis to verify nominal resolving power, polarimetric performance and S/N. Preliminary LSD profiles are extracted using our database of generic hot star line masks to perform an initial magnetic field diagnosis and further quality assurance. Finally, each ESPaDOnS spectrum will be processed by the MiMeS Massive Stars Pipeline (MSP; currently in production) to determine a variety of critical physical data for each observed target, in addition to the precision magnetic field diagnosis: effective temperature, surface gravity, mass, radius, age, variability characteristics, projected rotational velocity, radial velocity and binarity, and mass loss rate. These meta-data, in addition to the reduced high-quality spectra, will be uploaded for publication to the MagIcS Legacy Database\footnote{The MiMeS Project is undertaken within the context of the broader MagIcS (Magnetic Investigations of various Classes of Stars) collaboration, www.ast.obs-mip.fr/users/donati/magics).}.

\section{First results}

MiMeS operations began in August 2008, and nearly one semester of observations has been acquired at the time of writing. This corresponds to approximately 70 hours of observation, during which more than 200 polarised spectra were acquired for about 50 MiMeS targets. Further details about the first results of the MiMeS Project are reported by Grunhut et al. (these proceedings).

\begin{acknowledgments}
The MiMeS Large Program is supported by both Canada and France, and was one of 4 such programs selected in early 2008 as a result of an extensive international expert peer review of many competing proposals.

Based on observations obtained at the Canada-France-Hawaii Telescope (CFHT) which is operated by the National Research Council of Canada, the Institut National des Sciences de l'Univers of the Centre National de la Recherche Scientifique of France, and the University of Hawaii.

The MiMeS Data Access Pages are powered by software developed by the CADC, and contains data and meta-data provided by the CFHT Telescope.
\end{acknowledgments}




\end{document}